\documentclass[12pt,preprint]{emulateapj} 
\def\be{\begin{equation}}
\def\ee{\end{equation}}

\def\lsim{\lower 2pt \hbox{$\, \buildrel {\scriptstyle <}\over
         {\scriptstyle \sim}\,$}}
\newcommand\gsim{\buildrel > \over \sim}
\begin{document}
\newcommand{\figureout}[2]{ \figcaption[#1]{#2} }       

\title{Discovery of Pulsed $\gamma$-Rays from the Young Radio Pulsar PSR J1028-5819 
with the {\it Fermi} Large Area Telescope}

\author{
A.~A.~Abdo\altaffilmark{1,2}, 
M.~Ackermann\altaffilmark{3}, 
W.~B.~Atwood\altaffilmark{4}, 
L.~Baldini\altaffilmark{5}, 
J.~Ballet\altaffilmark{6}, 
G.~Barbiellini\altaffilmark{7,8}, 
M.~G.~Baring\altaffilmark{9}, 
D.~Bastieri\altaffilmark{10,11}, 
B.~M.~Baughman\altaffilmark{12}, 
K.~Bechtol\altaffilmark{3}
% ...
}
\altaffiltext{1}{National Research Council Research Associate}
\altaffiltext{2}{Space Science Division, Naval Research Laboratory, Washington, DC 20375}
\altaffiltext{3}{W. W. Hansen Experimental Physics Laboratory, Kavli Institute for Particle Astrophysics and Cosmology, Department of Physics and Stanford Linear Accelerator Center, Stanford University, Stanford, CA 94305}
\altaffiltext{4}{Santa Cruz Institute for Particle Physics, Department of Physics and Department of Astronomy and Astrophysics, University of California at Santa Cruz, Santa Cruz, CA 95064}
\altaffiltext{5}{Istituto Nazionale di Fisica Nucleare, Sezione di Pisa, I-56127 Pisa, Italy}
\altaffiltext{6}{Laboratoire AIM, CEA-IRFU/CNRS/Universit\'e Paris Diderot, Service d'Astrophysique, CEA Saclay, 91191 Gif sur Yvette, France}
\altaffiltext{7}{Istituto Nazionale di Fisica Nucleare, Sezione di Trieste, I-34127 Trieste, Italy}
\altaffiltext{8}{Dipartimento di Fisica, Universit\`a di Trieste, I-34127 Trieste, Italy}
\altaffiltext{9}{Rice University, Department of Physics and Astronomy, MS-108, P. O. Box 1892, Houston, TX 77251, USA}
\altaffiltext{10}{Istituto Nazionale di Fisica Nucleare, Sezione di Padova, I-35131 Padova, Italy}
\altaffiltext{11}{Dipartimento di Fisica ``G. Galilei", Universit\`a di Padova, I-35131 Padova, Italy}
\altaffiltext{12}{Department of Physics, Center for Cosmology and Astro-Particle Physics, The Ohio State University, Columbus, OH 43210}
% ...


\begin{abstract}
Radio pulsar PSR J1028-5819 was recently discovered in a high-frequency search (at 3.1 GHz)
in the error circle of the EGRET source 3EG J1027-5817.   The spin-down power of this young pulsar
is great enough to make it very likely the counterpart for the EGRET source.  We report here the 
discovery of
$\gamma$-ray pulsations from PSR J1028-5819 in early observations by the Large Area Telescope (LAT) 
on the {\it Fermi} Gamma-Ray Space Telescope.  
The $\gamma$-ray light curve shows two sharp peaks having phase 
separation of $0.460 \pm 0.004$, trailing the very narrow radio pulse by $0.200 \pm 0.003$ in phase, 
very similar to that of other known $\gamma$-ray pulsars.  The measured $\gamma$-ray flux  
gives an efficiency for the pulsar of $\sim 10-20\%$ (for outer magnetosphere beam models). 
No evidence of a surrounding pulsar wind nebula is seen in the current {\it Fermi} data but limits on associated
emission are weak because the source lies in a crowded region with high background emission.
However, the improved angular resolution afforded by the LAT enables the disentanglement of the previous 
COS-B and EGRET source detections into at least two distinct sources, one of which is now identified as PSR J1028-5819. 
\end{abstract} 

\keywords{pulsars: general --- stars: neutron}

\pagebreak
  
\section{Introduction}

One of the most intriguing legacies of the Energetic Gamma-Ray Experiment Telescope (EGRET) on the Compton 
Gamma-Ray Observatory ({\it CGRO}) was the group of 150 sources (Hartman et al. 1999) 
that could not be firmly identified with any
known counterparts.  The majority of these (about 100) are in a population clustered toward the Galactic plane, and 
many were presumed to be $\gamma$-ray pulsars.  
Indeed, a number of radio pulsars were discovered to lie in or near EGRET 
unidentified source error boxes after the end
of CGRO observations in 2000, both in large surveys such as the Parkes Multibeam Survey (Kramer et al. 2003) or 
in deep observations in some of the EGRET error boxes (e.g. Roberts et al. 2002, Halpern et al. 2001).  
Unfortunately, searches for pulsations at the radio periods in the
EGRET archival data were not feasible, given the small number of $\gamma$-ray photons in these sources  
and the difficulty of predicting pulsar phase in the presence of rotational instabilities (also known as timing noise).

The situation has improved dramatically with the recent launch of {\it Fermi} on June 11, 2008, which has been operating 
successfully through early calibration and now in sky survey mode.  The {\it Fermi} LAT has a sensitivity that is 
more than an order of magnitude that of EGRET, and it is already possible to better perform searches for $\gamma$-ray
pulsations in many of the EGRET sources.  This Letter reports results on one such search in the error circle of the EGRET source 3EG J1027-5817, at the location of the radio pulsar PSR J1028-5819, 
discovered (Keith et al. 2008) just a few months prior to the launch of 
{\it Fermi} as part of a search of three EGRET sources at high frequency.  PSR J1028-5819 is a young pulsar,
with period $P = 91.4$ ms, period derivative $\dot P = 1.61 \times 10^{-14}\,\rm s\,s^{-1}$ and characteristic age of $9.21 \times 10^{4}$ yr.  The derived spin-down power, $\dot E_{\rm sd} = 8.43 \times 10^{35}
\,\rm erg\,s^{-1}$ combined with its dispersion measure-derived distance of 2.3 kpc makes it a plausible counterpart for the EGRET source with flux of $6.6 \pm 0.7 \times 10^{-7}\,\rm ph\, cm^{-2}\,s^{-1}$ ($E > 100$ MeV).  
The radio pulse profile is extremely narrow, consisting of two highly
linearly polarized components.  The full width is only 560 $\mu$s,
giving it the smallest duty cycle of any known pulsar and an order
of magnitude smaller duty cycle than other young pulsars
(e.g. Weltevrede \& Johnston 2008). {\it Fermi} LAT has now discovered the $\gamma$-ray pulsations 
from PSR J1028-5819, and we can thus confirm that some of the photons attributed to 3EG J1027-5817 originate from PSR J1028-5819.

\section{Observations}

{\it Fermi} was launched into low-Earth orbit and, after a six-week commissioning phase, began nominal
sky-survey observations on August 11, 2008.  The LAT, the main instrument on {\it Fermi}, is a pair-production
telescope (Atwood et al. 2009) sensitive to $\gamma$ rays from 20 MeV to at least 300 GeV with on-axis 
effective area $> 1$ GeV of $\sim 8000\, \rm cm^2$, exceeding that of EGRET by a factor of about five.  
It has a large field-of-view of 2.4 sr and in survey mode observes the entire sky every 3 hr.
We report here on observations using data collected during the
initial 35 days in on-orbit verification that included sky-survey tuning and pointed-mode tuning on Vela, 
from June 30 - August 3, 2008, as well as the initial 15 weeks of sky survey, from August 3 - November 16, 2008.
The Diffuse class events (LAT event class having the tightest background rejection)
from these periods total 6014 photons with energy $> 100$ MeV, within a radius of $1\fdg 5$ surrounding 
PSR J1028-5819.  We excluded periods when the pulsar was viewed at 
zenith angle $> 105^\circ$ to the detector axis where the Earth's albedo photons gave excessive background contamination.

Radio observations of PSR J1028$-$5819 were carried out at the
Parkes 64-m radio telescope at frequencies near 1.4 and 3.1~GHz
with typical durations of $\sim$5~min. Timing observations
commenced on 2008 April 7, shortly after the discovery of the radio pulsar.
Since then 24 independent timing measurements have been made with
typical uncertainties in the times-of-arrival of $\sim$30~$\mu$s or better.
The fit to the timing points was carried out using TEMPO2 (Hobbs et al. 2006); the
position of the pulsar was held fixed at the value given in Keith
et al. (2008) and the rotation frequency and frequency derivative
were fit. The dominant contribution to the resulting residual of 270~$\mu$s
is timing noise intrinsic to the pulsar.  Note that the timing solution
used to derive the narrow radio profiles contains an extra fit parameter
that is not included in the {\it Fermi} software tools used for the $\gamma$-ray profile.  
Including this additional parameter leads to at most a phase shift of 0.01 over the dataset 
used for the $\gamma$-ray pulsation analysis.

\section{Results}

For the detection of pulsations, Diffuse class events with energy $> 100$ MeV and within a 
radius of $1\fdg 5$ of the radio position were corrected to the Solar-System barycenter using the JPL DE405
Solar System ephemeris
and folded with the radio period using the Parkes ephemeris.  The {\it Fermi} LAT timing is derived from a GPS clock
on the satellite and photons are timestamped to an accuracy better than 300 ns.  
The LAT software tools for pulsars have been shown to be accurate to a few $\mu$s for isolated pulsars (Smith et al 2008).
We detect $\gamma$-ray pulsations at the radio period with chance probability $2 \times 10^{-27}$ using a $Z_n$-test with
2 harmonics (DeJager et al. 1989).  Within 2 weeks from the onset of data collection, 
a $Z_n$ significance of 3.5 $\sigma$ was found for pulsations, and this was improved to a better than 10 $\sigma$
pulsed signal with less than 4 months of LAT data.
The $\gamma$-ray pulsations at the same period were independently found 
by a blind search in $P$ and $\dot P$ using a time-differencing technique (Atwood et al. 2006, Ziegler et al. 2008).

The LAT has an angular resolution with a dependence on reconstructed event energy $E$ of $E^{-0.75}$, with a 68\% 
containment radius of $\sim 0\fdg 5$ at 1 GeV for near on axis events which convert in one of the first 12 layers 
of the tracker (the thin section) and that increases with incidence angle as detailed in Atwood et al (2009).  Events 
converting in one of the last four layers of the tracker (the thick section) have a 68\% containment radius which 
is $\sim 2$ times that of the thin section.  In order to explore the energy dependence of the light curve, the event 
selection was refined to be $\theta_c(E/100\,{\rm MeV})^{-0.75}$ , where $\theta_c = 3^\circ$ for thin events and 
$\theta_c = 4\fdg 1$ for thick events are the containment radii at 100 MeV chosen to maximize the detection 
significance with this energy dependent cut. 
The histogram of folded counts at energies 0.1 - 13 GeV is shown in Figure 1.  
The $\gamma$-ray light curve shows two strong peaks, P1 at phase $0.200 \pm 0.003$ and P2 at phase $0.661 \pm 0.002$, 
where phase 0 is defined by the dedispersed radio pulse. The phase separation of P1 and P2 is $0.460 \pm 0.004$.
The peaks are fairly narrow, with Lorentzian FWHM for P1 of 
$0.040 \pm 0.011$  and for P2 of $0.035 \pm 0.007$.
This light curve is very similar to that of Vela, as seen with EGRET (Kanbach et al. 1994) 
and now with {\it Fermi} (Abdo et al. 2009a).  
Figure 2 shows light curves in four different
energy ranges, $100 - 300$ MeV, 300 MeV - 1 GeV, 1-3 GeV and $> 3$ GeV that do not exhibit significant evolution in 
shape.  In particular, we have measured the widths of the peaks as a function of energy and for P1 there is a deviation at only the 2.19 $\sigma$ level and for P2 at the 1.63 $\sigma$ level.  There is also no evidence for significant evolution of the P1/P2 ratio.  A fit of the P1/P2 values to a constant function of energy gives a $\chi^2$ per degree of 
freedom of 0.54, which is consistent with no variation.  This constrasts to the  
decrease in the P1/P2 ratio with energy seen in the Crab, Vela, Geminga and B1951+32 pulsars by EGRET 
(Thompson 2004), and is now confirmed in the Vela pulsar by {\it Fermi} (Abdo et al. 2009a).   
We have measured the pulsed significance of PSR J1028+5819 as a function of increasing energy, successively raising the low energy threshold.  For all events above 4 GeV (with the energy dependent cut) the pulsed detection is 3.5 $\sigma$, but above 5 GeV it is only 1.8 $\sigma$, indicating that the maximum energy of pulsations is around 4 GeV.

The LAT point source 0FGL J1028.6-5817 from the {\it Fermi} LAT bright source list 
(Abdo et al. 2009b) corresponding to PSR J1028-5819 is located at
(RA,DEC)=(157.166,-58.292) with a 95\% confidence level radius of 0.079 degrees.
There are two other LAT point sources nearby, 0FGL J1024.0-5754 and 
0FGL J1018.2-5858, $0\fdg 73$ and $1\fdg 52$ away respectively.
The COS-B source 2CG 284-00 (Swanenburg et al. 1981) 
was apparently made up of contributions from all three LAT sources while the EGRET source 
3EG J1027-5817 has now been resolved by the {\it Fermi} LAT into contributions from the two sources, 
0FGL J1028.6-5817 and 0FGL J1024.0-5754.  
The LAT source associated with PSR J1028-5819 is therefore somewhat confused 
with the nearby point sources and contains significant photon flux from 0FGL J1024.0-5754. 
This adds to the unpulsed background and must be
taken into account in computing the phase-averaged flux from the pulsar.
A $Swift$ XRT (Burrows et al. 2005) observation on November 23-24, 2008
(9.6 ks cleaned exposure) yielded a 4.1$\sigma$ detection of an X-ray
source that we tentatively associate with the pulsar. Although the X-ray
source is positionally coincident with the radio pulsar, it is poorly
localized ($\sim$20$''$ total extent) and is faint, with a net count
rate (0.3--10 keV) of 2.3 $\times$ 10$^{-3}$ cts/sec which corresponds
to an absorbed (unabsorbed) flux of 1.5 (3.4) $\times$ 10$^{-13}$ erg
cm$^{-2}$ s$^{-1}$, assuming a power-law with $\Gamma$=2 and $N_{\rm H}$
= 1.59 $\times$ 10$^{22}$ cm$^{-2}$.

To obtain the phase-averaged flux of the pulsar, we have performed a maximum likelihood spectral 
analysis using the LAT
tool gtlike$\footnote{http://fermi.gsfc.nasa.gov/ssc/data/analysis/SAE\_overview.html}$ on counts 
within a radius of $15^\circ$ from the LAT source position.  
The pulsar spectrum was fit with a power law with an exponential cutoff, giving an 
index $1.22 \pm 0.2 \pm 0.12$ and cutoff energy of $2.5 \pm 0.6 \pm 0.5$ GeV 
(the first errors are statistical and the second are systematic).  With the current statistics in this 
crowded region, a differentiation between a simple and a hyper-exponential cutoff is not yet possible and, 
therefore, was not attempted.
From this fit, we obtain an integral photon flux at $0.1 - 30$ GeV of $1.62 \pm 0.27 \pm 0.32 \times 10^{-7}\,\rm ph\,cm^{-2}\,s^{-1}$ 
and integral energy flux of $1.78 \pm 0.15 \pm 0.35 \times 10^{-10}\,\rm erg\,cm^{-2}\,s^{-1}$.  
This flux is a quarter of the EGRET source flux, which apparently included contributions from both 
PSR J1028-5819 and its 
neighboring source 0FGL J1024.0-5754.  
In addition to the pulsar, the diffuse $\gamma$-ray emission from the Milky Way was modeled, 
while the extragalactic diffuse emission plus instrumental residual background
and other LAT point sources within the region of interest were fit with power law spectra.
The LAT source 0FGL J1024.0-5754 which likely contributed to the EGRET source was fit with an integrated
flux from $0.1 - 30$ GeV of $3.0 \times 10^{-7}\,\rm ph\,cm^{-2}\,s^{-1}$.  
The flux from this source plus that from PSR J1028-5819 add 
up to $\sim 4.8 \times 10^{-7}\,\rm ph\,cm^{-2}\,s^{-1}$ which is consistent,
within statistical and systematic uncertainties, with the flux $6.6 \pm 0.7 \times 10^{-7}\,\rm ph\, cm^{-2}\,s^{-1}$ 
of 3EG J1027-5817.  The 3EG source flux was derived by assuming a power law with index 2 (Hartman et al. 1999) 
which would have preferentially weighted the spectrum  
with a larger PSF, adding more counts from the background at low energy to produce a higher flux.  
The LAT source 0FGL J1018.2-5858, which likely contributed to the flux of the 2CG source, 
yielded an integrated flux at $0.1 - 30$ GeV of $5.41 \times 10^{-7}\,\rm ph\,cm^{-2}\,s^{-1}$ 
from a power-law fit.  Adding the flux from  PSR J1028-5819, 0FGL J1024.0-5754 and 0FGL J1018.2-5858 then 
could account for the 2CG flux of $\sim 2.7 \times 10^{-6}\,\rm ph\,cm^{-2}\,s^{-1}$.

\section{Discussion}

The detection of pulsed $\gamma$ rays from the recently discovered pulsar PSR J1028-5819 
within the first few weeks of the {\it Fermi} mission confirms the promise that the {\it Fermi} LAT
will be an important instrument for pulsar studies.  The $\gamma$-ray pulses of PSR J1028-5819 cover a 
wide phase range and neither of the peaks is aligned with the very narrow radio pulse.
This strongly suggests that the $\gamma$-ray beam covers a large fraction of solid angle of the sky and favors
its interpretation in outer magnetosphere models such as the outer gap 
(OG) (Cheng, Ho \& Ruderman 1986, Romani \& Yadigaroglu 1995) or slot gap (SG) (Muslimov \& Harding 2004).  
Furthermore, the maximum observed energy of pulsations, 
$\epsilon_{max} \simeq 4$ GeV must lie below any $\gamma$-B pair production turnover threshold, 
thereby providing a lower bound to the altitude of emission, even though we have not been able to rule out the 
hyper-exponential spectral cutoff expected for pair production attenuation. 
Using a standard polar cap model estimate for the
minimum emission height of $r\gtrsim (\epsilon_{max} B_{12}/1.76\, {\rm GeV})^{2/7}\, P^{-1/7}\, R_\ast$ (inverting 
Eq. [1] of Baring 2004), for a surface polar field strength of $B_{12} = B/10^{12}$ G and neutron star radius $R_\ast$, 
the PSR J1028-5819 spin-down parameters 
($P=0.0914$s, $B_{12}=1.015$) yield $r\gtrsim 1.8 R_\ast$. This bound precludes emission very near the stellar 
surface.

In order to estimate the $\gamma$-ray
efficiency of a pulsar, one needs to know the total luminosity radiated, $L_{\gamma}^{tot} = 4\pi f(\alpha, \zeta_E) \Phi_{obs}
d^2$, where $f$ is a correction factor that contains information about the beaming geometry, $\Phi_{obs}$ is the 
observed phase-averaged energy flux, and $d$ is the distance to the source.  The 
factor $f(\alpha, \zeta_E)$ is a function of the pulsar magnetic inclination angle $\alpha$ and the 
viewing angle to the rotation axis, $\zeta_E$, is very model sensitive and can be computed for any particular model 
geometry by,
\be \label{f}
f(\alpha, \zeta_E) = {\int F(\alpha, \zeta, \phi) d\Omega \over 4\pi \Phi_{obs}}
\ee
(Watters et al. 2008). 
Here, $F(\alpha, \zeta, \phi)$ is the radiated flux from a pulsar as a function of inclination angle, 
viewing angle and rotation phase, $\phi$.
The factor $f(\alpha, \zeta_E)$ is very important because it quantifies the amount of the emission 
we may be missing with our limited sweep over the pulsar beam. 
For polar cap models where the emission originates within several stellar radii of the neutron star 
surface, the effective emission solid angle is small and thus the factor $f \ll 1$.  For outer magnetosphere models,
the traditional solid angle measure is not appropriate since the emission is radiated over a large
fraction of $4\pi$.  The beaming correction factor $f$ must therefore be
computed numerically using Eqn (\ref{f}) and it is found that $f \gsim 1$ for both OG and SG models (Watters et al. 2008).  
For PSR J1028-5819, we obtain a total luminosity of $L_{\gamma}^{tot} = 1.1 \times 10^{35}\,f\,\rm erg\,s^{-1}$ 
from the observed energy flux and source distance, $d = 2.3$ kpc.  The $\gamma$-ray efficiency is thus $\eta_{\gamma} = L_{\gamma}^{tot}/\dot E_{sd} = 0.13\,f/I_{45}$, where $I_{45} = 
I/10^{45}\,\rm g\,cm^2$ is the neutron star moment of inertia.  Since the distance could have as much as a $40 \%$ 
error from fluctuation in the free electron density (Brisken et al. 2002), there is an uncertainty of about a factor of 2 in 
the derived luminosity and the efficiency. 
If the outer magnetosphere interpretation of the pulsed $\gamma$ rays from PSR J1028-5819  and other young pulsars is correct, 
then these pulsars have efficiencies that
are larger by about an order of magnitude than would be deduced using the previously standard 1 sr
solid angle. 

The full geometry of PSR J1028-5819 is not easily determined from the radio data alone because the
narrowness of the pulse make it difficult to derive good solutions from polarization position angle variation using the 
rotating vector model. But the very narrow pulse does argue for a large $\alpha$ (in agreement with the constraints below).  In the outer-magnetosphere geometry now favored by the $\gamma$-ray emission, 
one can derive constraints on $\alpha$ and $\zeta$ from the measured separation of the $\gamma$-ray peaks
and $\gamma$-ray efficiency alone.  Using the $\gamma$-ray light curve `Atlas' of Watters et al. (2008), we
estimate an allowed range of $\alpha \sim 70^\circ - 90^\circ$, $\zeta \sim 75^\circ - 80^\circ$ and $f \sim 1.1$ for the 
OG model and $\alpha \sim 65^\circ - 80^\circ$, $\zeta \sim 60^\circ - 80^\circ$ and $f \sim 0.9 - 1.0$ for the 
two-pole caustic (TPC) or SG models.  Both of these models therefore have a good range for viable solutions.  However,
the above estimates assume the $\gamma$-ray efficiency relation 
$\eta_{\gamma} \simeq (10^{33}\,{\rm erg\,s^{-1}}/\dot E_{sd})^{1/2}$ which gives $\eta_{\gamma} = 0.03$, a factor of 4 smaller
than that derived for this pulsar from its measured luminosity.  Using our derived $\eta_{\gamma} = 0.13$, the range
allowed for $\alpha$ shrinks to $\sim 80^\circ - 90^\circ$ for the OG model, but remains about the same for the TPC model.
Detection of a pulsar wind nebula (which can give an estimate of $\zeta_E$) or radio polarization measurement (which might 
give an $\alpha$ estimate) would help further constrain the model.  
The phase lag of the first $\gamma$-ray peak relative to the radio pulse can be explained in either
OG or SG (or generally TPC, Dyks \& Rudak 2003) models, where the radio phase crossing associated 
with a magnetic pole occurs before the first high-energy emission caustic formed at high altitude.

If many of the other young radio pulsars discovered in EGRET error boxes are similar to PSR J1028-5819 and if 
the $\gamma$-ray beams of young pulsars can be seen from a wide range of viewing angles, then 
{\it Fermi} searches for their pulsed $\gamma$ rays promise to be very fruitful.  The idea of wide $\gamma$-ray beams 
recently gained strong support from the {\it Fermi} LAT discovery of a radio quiet pulsar in the young supernova remnant 
CTA 1 (Abdo et al. 2008) whose pulsations have been detected only in $\gamma$ rays, a blind search result secured 
promptly during the {\it Fermi} commissioning phase.  
The discovery of $\gamma$-ray pulsations from a young radio pulsar coincident with an EGRET source, together with the discovery of $\gamma$-ray pulsations from CTA 1, confirms expectations from before the {\it Fermi} launch that many $\gamma$-ray pulsars remain to be discovered. Even more exciting are the prospects that the all-sky survey of the {\it Fermi} LAT will identify many new sources, and that searches in both the $\gamma$-ray and radio bands will uncover an entirely new population of pulsars.

\vskip 0.5cm

The $Fermi$ LAT Collaboration acknowledges generous ongoing support from a number of agencies and institutes that have supported both the development and the operation of the LAT as well as scientific data analysis.  These include the National Aeronautics and Space Administration and the Department of Energy in the United States, the Commissariat \`a l'Energie Atomique and the Centre National de la Recherche Scientifique / Institut National de Physique Nucl\'eaire et de Physique des Particules in France, the Agenzia Spaziale Italiana and the Istituto Nazionale di Fisica Nucleare in Italy, the Ministry of Education, Culture, Sports, Science and Technology (MEXT), High Energy Accelerator Research Organization (KEK) and Japan Aerospace Exploration Agency (JAXA) in Japan, and the K.~A.~Wallenberg Foundation, the Swedish Research Council and the Swedish National Space Board in Sweden.

Additional support for science analysis during the operations phase from the following agencies is also gratefully acknowledged: the Istituto Nazionale di Astrofisica in Italy and the K.~A.~Wallenberg Foundation in Sweden for providing a grant in support of a Royal Swedish Academy of Sciences Research fellowship for JC.

The Parkes radio telescope is part of the Australia Telescope which is funded by the Commonwealth of Australia for operation as a National Facility managed by the CSIRO.

\vskip 0.2cm
\noindent
\flushleft{Contact authors: Tyrel Johnson (Tyrel.J.Johnson@nasa.gov) \& Alice Harding (Alice.K.Harding@nasa.gov)}

\acknowledgments  

\newpage
\begin{figure}
\includegraphics[width=160mm]{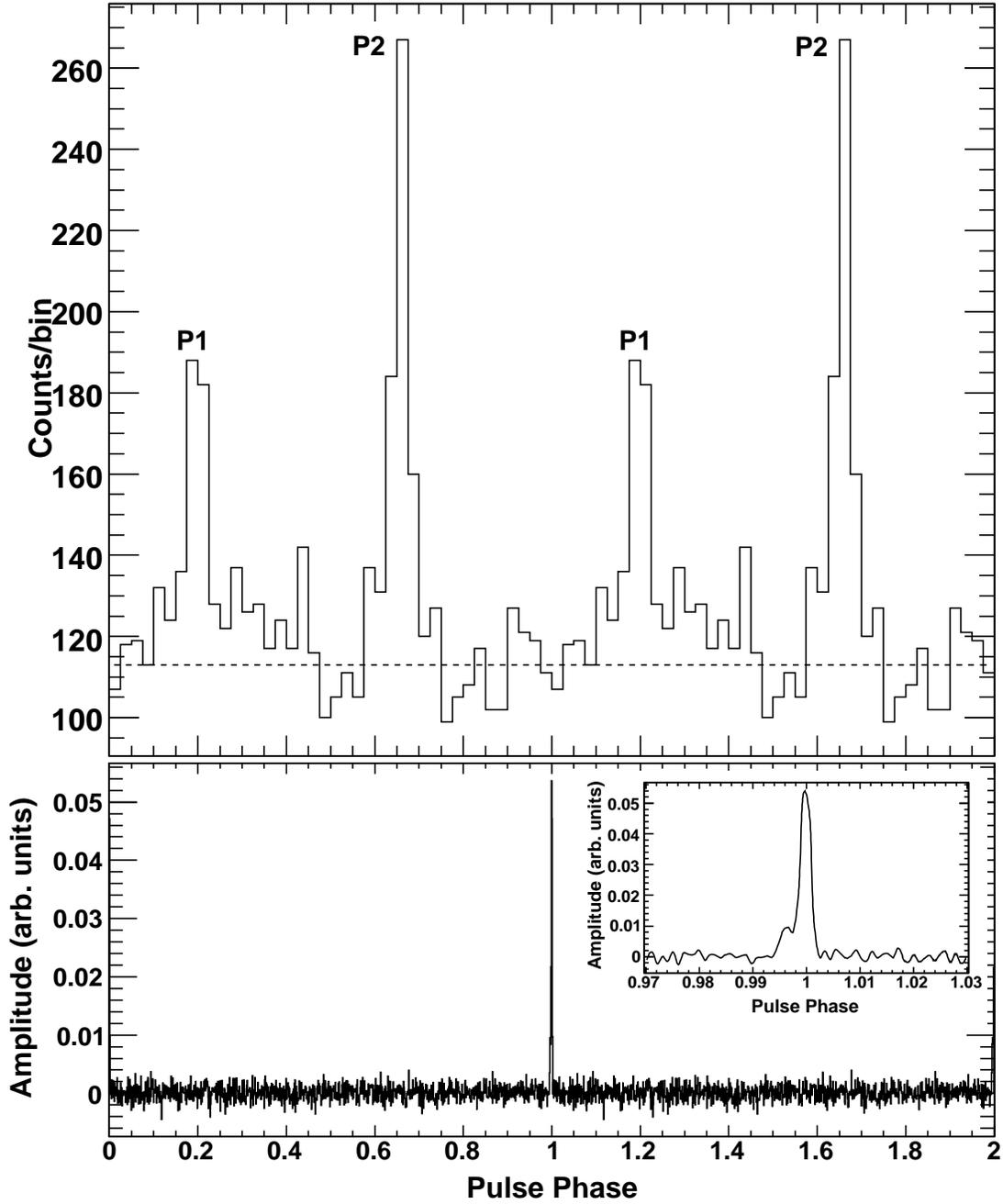}
\vskip 1.0cm
\caption{Light curve of PSR J1028-5819 in the (0.1 - 13 GeV) band in 40 constant-width bins and shown over 
two pulse periods with the 1.4 GHz radio pulse profile plotted below.  The horizontal dashed line shows the estimated
background level from the off-pulse region at phase 0.8 and 1.0.
The inset shows the radio pulse in the phase range 0.97 - 1.03, with the main peak at phase 1.0 and preceded by the smaller, secondary peak at phase $\sim 0.996$.}
\end{figure}

\newpage
\begin{figure}
\hskip -1.0cm
\includegraphics[width=200mm]{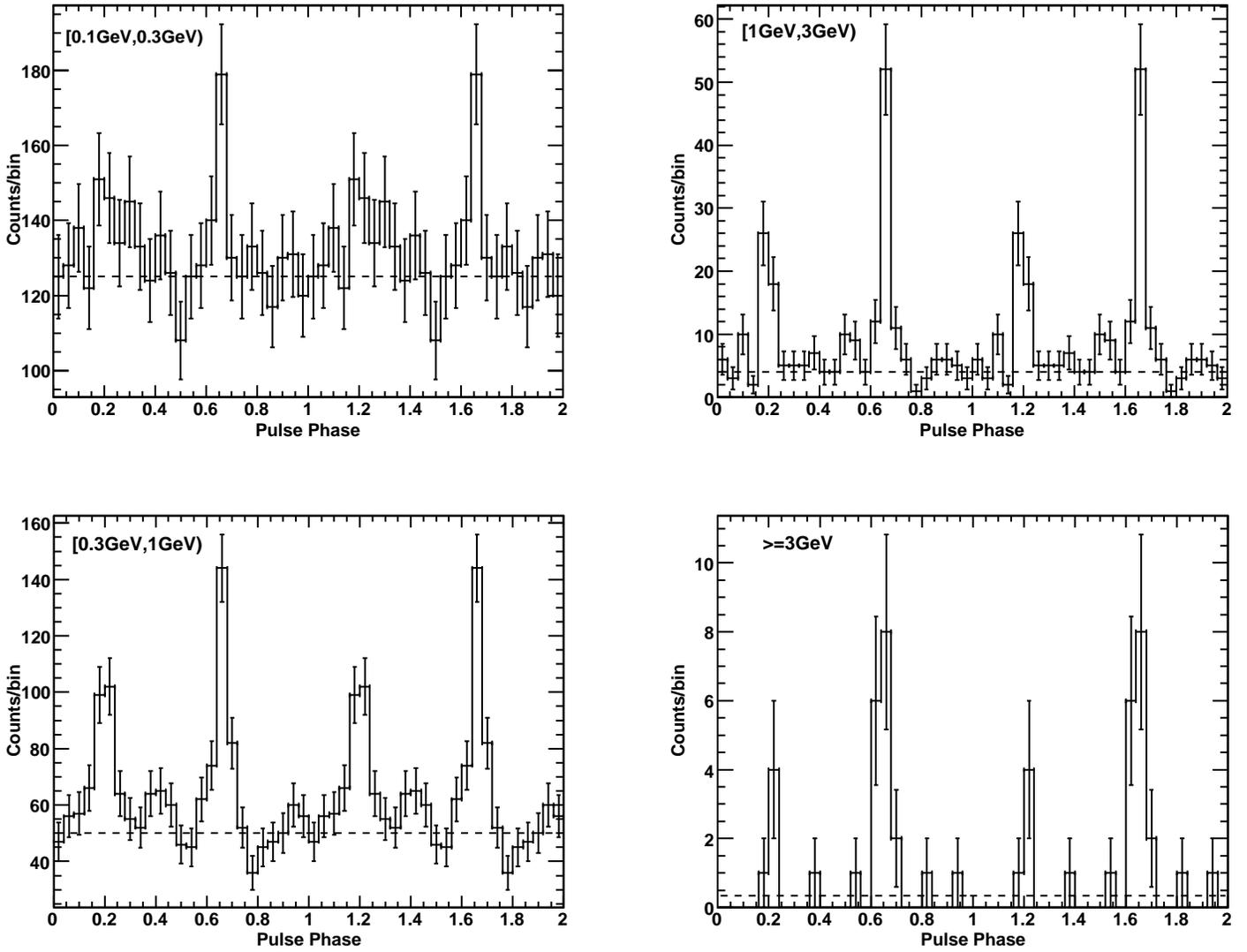}
\vskip 1.0cm
\caption{Light curves of PSR J1028-5819 in four different energy bands (labeled) in constant width bins of size 0.04 in phase.
The horizontal dashed lines show the estimated background level from the off-pulse region at phase 0.8 and 1.0.
}
\end{figure}
\end{document}